\documentclass[twocolumn]{aastex631}

\newcommand{\msun}{M_\odot}
\newcommand{\rsun}{R_\odot}

\newcommand{\ace}{$\alpha_\mathrm{CE}$}

\usepackage{CJK}
\usepackage{acronym}
\usepackage{xcolor}
\usepackage{ulem}

\acrodef{ce}[CE]{common-envelope}
\acrodef{rsg}[RSG]{red supergiant}
\acrodef{rg}[RG]{red giant}

\begin{document}
\begin{CJK*}{UTF8}{ipxm}
\title{A two-stage formalism for common-envelope phases of massive stars}

\author[0000-0002-8032-8174]{Ryosuke Hirai (平井遼介)}
\affiliation{School of Physics and Astronomy, Monash University, VIC 3800, Australia}
\affiliation{OzGrav: The Australian Research Council Centre of Excellence for Gravitational Wave Discovery, Clayton, VIC 3800, Australia}

\author[0000-0002-6134-8946]{Ilya Mandel}
\affiliation{School of Physics and Astronomy, Monash University, VIC 3800, Australia}
\affiliation{OzGrav: The Australian Research Council Centre of Excellence for Gravitational Wave Discovery, Clayton, VIC 3800, Australia}



\begin{abstract}
We propose a new simple formalism to predict the orbital separations after common-envelope phases with massive star donors. We focus on the fact that massive red supergiants tend to have a sizeable radiative layer between the dense helium core and the convective envelope. Our formalism treats the common-envelope phase in two stages: dynamical in-spiral through the outer convective envelope and thermal timescale mass transfer from the radiative intershell. With fiducial choices of parameters, the new formalism typically predicts much wider separations compared to the classical energy formalism. Moreover, our formalism predicts that final separations strongly depend on the donor evolutionary stage and companion mass. Our formalism provides a physically-motivated alternative option for population synthesis studies to treat common-envelope evolution. This treatment will impact on predictions for massive-star binaries, including gravitational-wave sources, X-ray binaries and stripped-envelope supernovae.
\end{abstract}



\section{Introduction} \label{sec:intro}

\Ac{ce} evolution remains one of the most uncertain phases of binary evolution \citep[see][for an overview on \ac{ce} evolution]{iva13,iva20}. In this phase, one of the stars in a binary system evolves to initiate unstable mass transfer, engulfing the companion star. The companion star will spiral in to orbit the core of the primary star inside a shared envelope, releasing a large amount of orbital energy. Part of the released energy can be transferred to the envelope to eject it. As a result, the binary is left with the companion star and the core of the primary in a very tight orbit.

The \ac{ce} phase was initially introduced to explain the origin of cataclysmic variables \citep[]{pac76}. Today, the idea is extended to higher-mass stars as well, and is considered responsible for explaining the origin of many other systems such as type Ia and stripped-envelope supernova progenitors, X-ray binaries and gravitational-wave sources, to name just a few. The main goal of \ac{ce} studies is to establish a way to predict the outcome, namely the post-\ac{ce} orbital separation, provided the pre-\ac{ce} binary properties. Despite the extensive efforts made via hydrodynamical modelling in 1D \citep[e.g.][]{iva15,cla17,fra19,kle21,mar21} and 3D \citep[e.g.][]{ras96,ric08,pas12,iva16,iac17,law20,gla21,mor21,lau22a,lau22b,gon22}, there is not yet a sufficient understanding of the physics to be able to provide accurate predictions. 

Many binary population synthesis studies instead adopt a simple parameterization to \ac{ce} phases based on energy balance (the so-called $\alpha$-formalism). This parameterization enables us to predict the outcome given an appropriate choice for the value of the parameter \ace. However, there is growing evidence both observationally and theoretically that the value for \ace{} may not be universal across different systems \citep[]{pol04,iac19}, questioning the suitability of the parameterization itself. In particular, there seems to be a qualitative discrepancy between low-mass and high-mass donors \citep[e.g.][]{wil22}, necessitating the use of different values of \ace{} for different mass regimes.

There is an alternative formalism based on angular momentum conservation, often called the $\gamma$-formalism \citep[]{nel00,nel05}. However, various studies point out that the ratio of the initial and final orbital angular momenta are so large that even a small change in model parameters will predict wildly different final separations \citep[]{web08,iva13,iva20}. Therefore, there seems to be little advantage of the $\gamma$ formalism over the $\alpha$ formalism as an outcome predictor.

In this letter, we propose a new framework to parameterize the \ac{ce} phase, with particular emphasis on massive star donors. In Section~\ref{sec:method}, we review the classical $\alpha$-formalism and then explain our new proposed framework. We demonstrate some examples in Section~\ref{sec:results} and conclude our study in Section~\ref{sec:conclusion}.

\section{Method}\label{sec:method}
\subsection{The classical energy formalism}

The orbital separations after \ac{ce} phases are often estimated with a simple framework based on energy conservation \citep[]{van76,tut79,web84}. The main idea is that the change in orbital energy of the binary is used to provide the energy required to unbind the envelope. This implicitly assumes that the orbital energy is fully efficiently converted into kinetic energy of the envelope, just enough to accelerate it to escape velocity. In reality the envelope can have outflow velocities above the escape velocity, or it can radiate away energy from the photosphere, causing it to ``waste'' part of the injected energy; on the other hand, there could be extra energy sources such as feedback from the accretor \citep[e.g.][]{fry96,sok04,cha18,lop19}, or magnetic fields \citep[]{ohl16}. To account for possible energy sources and sinks, a single efficiency factor is often introduced in the following form \citep[]{tut79,ibe85,dek90}:
\begin{equation}
 E_\mathrm{env}=\alpha_\mathrm{CE}\left(-\frac{GM_\mathrm{d}M_2}{2a_i}+\frac{GM_\mathrm{core}M_2}{2a_f}\right),\label{eq:alpha-formalism}
\end{equation}
where $E_\mathrm{env}$ is the binding energy of the envelope, $G$ is the gravitational constant, $M_\mathrm{d}$ is the mass of the pre-\ac{ce} donor, $M_2$ is the mass of the secondary, $M_\mathrm{core}$ is the mass of the donor's core, $a_i, a_f$ are the pre- and post-\ac{ce} orbital separations and \ace{} is the efficiency factor. In the absence of any extra energy sources, the efficiency should be $\alpha_\mathrm{CE}\leq1$.

The envelope binding energy is generally expressed as
\begin{equation}
 E_\mathrm{env}=\int^{M_\mathrm{core}}_{M_\mathrm{d}}\left(-\frac{Gm}{r}+\alpha_\mathrm{th}\epsilon\right)dm,
\end{equation}
where $m$ is the mass coordinate, $r=r(m)$ is the radius at a given mass coordinate and $\epsilon$ is the specific internal energy. The internal energy term contains several sub-components depending on the study. The envelope material contains a large amount of thermal energy which is required to maintain hydrostatic equilibrium. Once the equilibrium breaks down, the thermal energy can do work to accelerate the gas. For this reason, thermal energy is often considered as an extra energy source that can be used to eject the envelope and this is accounted for by applying that contribution to $E_\mathrm{env}$. Another commonly discussed energy source is the energy provided by recombination. This is a latent energy source that is only released when the gas expands and cools, but the total energy released can be significant especially in the outer parts of the envelope. To take into account this extra energy source, the ionization potentials for each ion are sometimes included in the internal energy term. In a recent study, we showed that while helium recombination may have a direct influence on the post-plunge separation, the influence of hydrogen recombination is limited \citep[]{lau22b}. In the following, we assume that $\epsilon$ includes the recombination energy of helium as well as the thermal energy of the gas and radiation. The coefficient $\alpha_\mathrm{th}$ is introduced to be able to tune the efficiency of initial internal energy in ejecting the envelope \citep[]{han95}.

Instead of directly using the binding energy itself in Eq.~(\ref{eq:alpha-formalism}), it is commonly parameterized as 
\begin{equation}
 E_\mathrm{env}=\frac{GM_\mathrm{d}M_\mathrm{env}}{\lambda R_\mathrm{d}},
\end{equation}
where $M_\mathrm{env}\equiv M_\mathrm{d}-M_\mathrm{core}$ is the envelope mass, $R_\mathrm{d}$ is the donor radius and $\lambda$ is an order unity parameter that expresses how centrally concentrated the density distribution of the envelope is. Population synthesis studies often employ either a fixed value for $\lambda$ or tabulated fits based on detailed stellar evolution calculations \citep[e.g.][]{dew00,xu10a,xu10b}.

Given all the pre-\ac{ce} binary properties ($E_\mathrm{env}, M_\mathrm{d}, M_2, a_i$) and the core mass of the primary ($M_\mathrm{core}$), the post-\ac{ce} separation ($a_f$) can be estimated from Eq.~(\ref{eq:alpha-formalism}). In many binary population synthesis studies, a fixed universal value is assumed for \ace{} to model the population of various observed systems. Calculations with various choices of \ace{} are carried out to quantify the uncertainties. However, both observations of post-\ac{ce} systems and 3D hydrodynamical simulations seem to suggest that the value of \ace{} is not universal and can depend on many factors such as donor mass, mass ratio, evolutionary stage, etc.~\citep[]{pol04,iac19}. In particular, massive donors have fundamentally different structures from low-mass donors (see next section) and hence are expected to behave qualitatively differently \citep[e.g.][]{fra19,wil22,lau22a}. There seems to be a building notion that corrections to the $\alpha$-formalism are required to explain various observations simultaneously \citep[e.g.][]{pol04,pol21,pol07}.

\subsection{A ``two-stage'' formalism}

The most obvious way to implement system-dependent \ac{ce} efficiencies is to create a functional form for \ace{} instead of a constant value. In principle, this can provide the correct post-\ac{ce} separations as long as the function is correct. However, having higher-order corrections to model parameters is often a sign that the model itself is not suitable. Here, we propose a new simple framework for treating \ac{ce} evolution that naturally behaves differently between low-mass and high-mass donors and also has a strong mass ratio dependence.

The main assumption that goes into the classical $\alpha$-formalism is that the whole \ac{ce} process is sufficiently shorter than the thermal timescale of the star so that it can be treated as an adiabatic process. This may be true for lower mass donors, as the thermal timescales of the envelopes are relatively long compared to the inspiral timescales. At higher masses, the thermal timescales are much shorter, and can become comparable to the inspiral timescale. It starts to become questionable whether the adiabatic treatment is adequate.

\begin{figure}
 \centering
 \includegraphics[width=\linewidth]{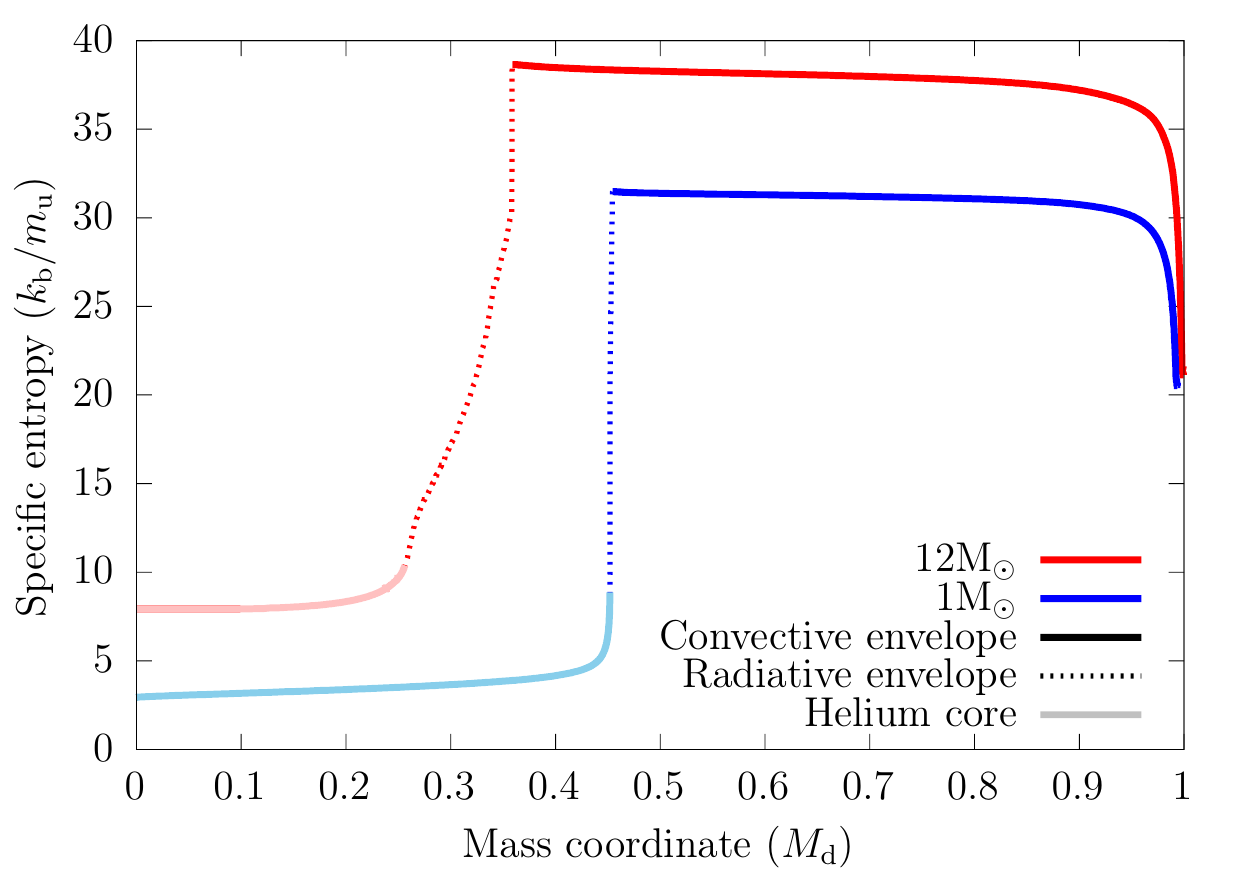}
 \caption{Entropy distribution in a $1~\msun$ \ac{rg} and $12~\msun$ \ac{rsg}. The $1~\msun$ \ac{rg} is chosen at a time when it has reached a radius of $100~\rsun$, and the $12~\msun$ \ac{rsg} is chosen when it has reached $600~\rsun$, just before the onset of core helium burning. The mass coordinate is normalized by the total mass. Solid parts of the curve indicate the convective layers while the dotted parts are radiative. The helium cores are colored with a lighter shade.\label{fig:entropy}}
\end{figure}

There is also a structural difference between low-mass \acp{rg} and high-mass \acp{rsg}. In Figure~\ref{fig:entropy}, we compare the entropy distributions inside a $1~\msun$ \ac{rg} and a $12~\msun$ \ac{rsg} computed with the public stellar evolution code MESA \citep[]{MESA1}. While for low-mass \acp{rg} the envelope is almost fully convective, high-mass \acp{rsg} have a radiative layer between the core and a convective outer layer. This intermediate radiative layer is much more tightly bound to the core, usually having an order of magnitude higher binding energy compared to the convective layer despite the contained mass being significantly less. Tabulated fits to the $\lambda$ parameter are therefore often quite small ($\lambda\sim\mathcal{O}(0.1)$) for higher mass stars \citep[]{dew00,xu10a,xu10b}.

An important feature of convective layers is that they have flat or slightly negative entropy gradients, while radiative layers have positive entropy gradients (Figure~\ref{fig:entropy}). This difference is crucial, as it determines the dynamical response of the stellar radius to mass loss \citep[]{sob97}. Roughly speaking, the stellar radius will contract in response to mass loss when the surface is radiative \citep[]{iva11}. This is because as the surface layers are removed, lower entropy material is exposed at the surface and therefore it cannot re-expand to its original size. On the other hand, convective envelopes can dynamically expand in response to mass loss due to the opposite argument. Thus, for a given mass ratio, mass transfer will be more unstable when the surface is convective and would likely lead to a \ac{ce} phase.

Once a \ac{ce} phase is initiated, the companion rapidly spirals in due to loss of corotation and dynamical friction with the envelope gas. According to 3D hydrodynamical simulations, the rapid plunge-in abruptly transitions into a slower spiral-in when the envelope is mostly ejected and corotation is achieved. For massive stars, the stalling can occur outside the radiative part of the envelope depending on the companion mass \citep[]{lau22a,lau22b}. 

After losing its convective envelope, the remaining radiative portion of the envelope does not dynamically expand but is still out of thermal equilibrium, so expands on a thermal timescale \citep[]{iva11,vig22}. This may initiate a more stable mass transfer to the companion star, which can further alter the orbit until most of the envelope is lost and the core detaches from its Roche lobe. At this point, energy conservation is no longer valid because by definition of the thermal timescale, the energy generated through nuclear burning during this phase becomes comparable to the binding energy. Therefore, it is more appropriate to model this phase with angular momentum conservation rather than energy balance arguments.

\begin{figure}
 \centering
 \includegraphics[width=\linewidth]{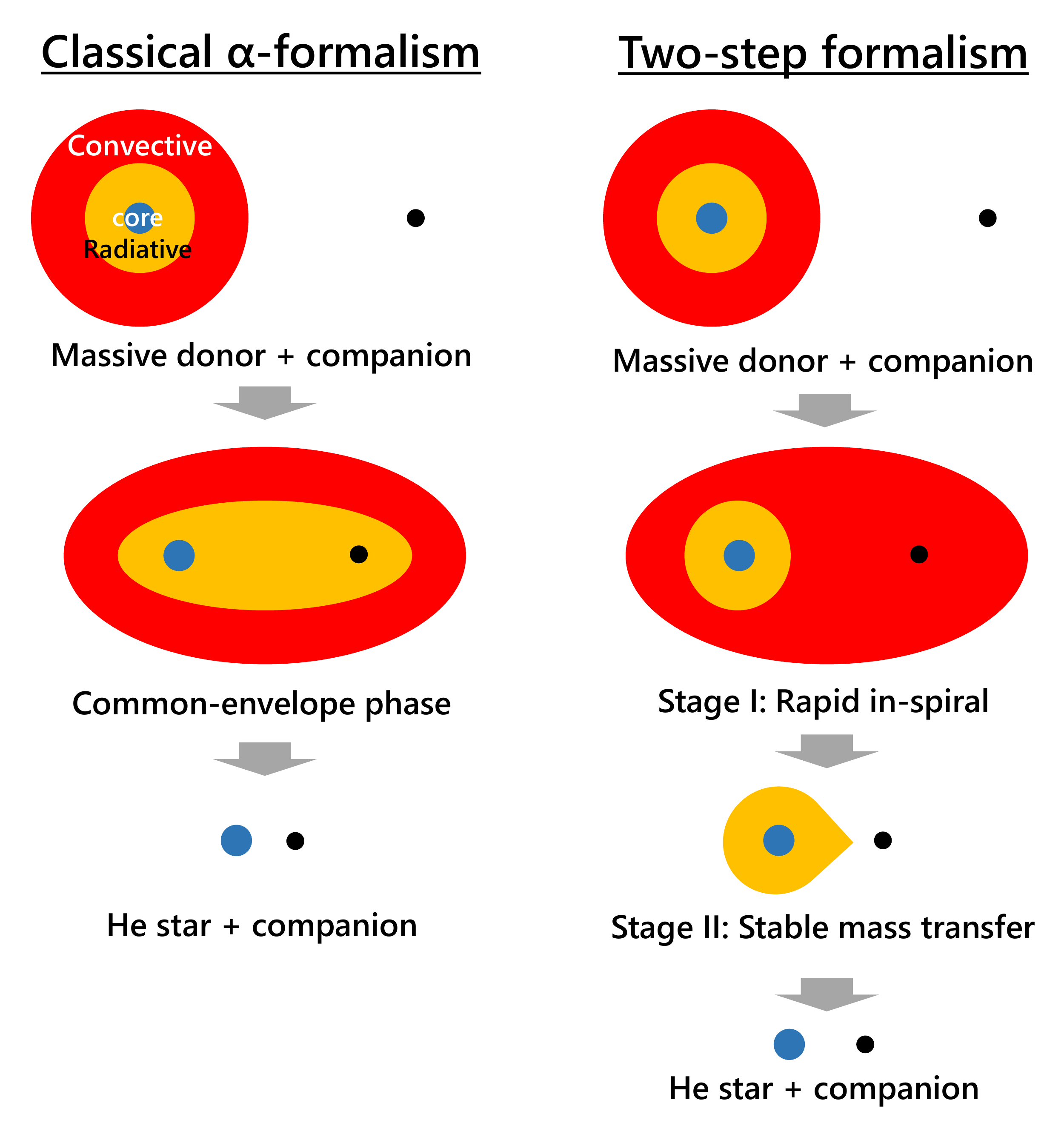}
 \caption{Schematic diagram of the traditional $\alpha$-formalism and the two-stage formalism we propose in this paper.\label{fig:sketch}}
\end{figure}

Building on these previous findings, we construct an alternative formalism for predicting \ac{ce} outcomes by splitting the whole process into two stages: I. the rapid spiral-in through the convective envelope, and II. the stable mass transfer of the radiative intershell. A schematic picture of our formalism compared to the classical $\alpha$-formalism is illustrated in Figure~\ref{fig:sketch}.
\\\textit{(Stage I)} We assume that the inspiral is much shorter than the thermal timescale and proceeds until the entire convective envelope is ejected. To estimate the orbital separation after stage I, we adopt the traditional $\alpha$-formalism while treating the radiative intershell as part of the core. Given the absence of extra energy sources on the dynamical timescale, we set $\alpha_\mathrm{CE}=1$ as a conservative estimate for the efficiency of stage I.
\\\textit{(Stage II)} We assume the mass transfer is stable and non-conservative, and the transferred matter takes away some angular momentum from the system. In general, the \ac{ce} phase is considered to occur when the companion mass is relatively small compared to the donor. Therefore, the companion should have a considerably longer thermal timescale compared to the donor and would not be able to accrete significant amounts of mass. In this paper, we assume for simplicity that the mass transfer is fully non-conservative, although some amount of accretion may in principle occur during this phase \citep[]{iva11}. We estimate the final orbital separation by solving for the system angular momentum evolution
\begin{equation}
 \frac{\dot{a}}{a}=\left(-2+\frac{M_\mathrm{d}}{M_\mathrm{d}+M_2}\right)\frac{\dot{M_\mathrm{d}}}{M_\mathrm{d}}+\frac{\dot{J}_\mathrm{orb}}{J_\mathrm{orb}},
\end{equation}
where $J_\mathrm{orb}$ is the orbital angular momentum \citep[]{pos14}. The value of $\dot{J}_\mathrm{orb}$ depends on the details of how the transferred material escapes the system. As a first attempt, we assume the matter is isotropically re-emitted from the accretor
\begin{equation}
 \frac{\dot{J}_\mathrm{orb}}{J_\mathrm{orb}}=\frac{\dot{M_\mathrm{d}}M_\mathrm{d}}{M_2(M_\mathrm{d}+M_2)}.
\end{equation}
Other choices could be made depending on the situation such as angular momentum loss from the $L_2$ point or from a circumbinary disk or some intermediate point similar to the $\gamma$-formalism as explored by \citet{mac18b}. With any choice of angular momentum loss  mode that is greater than isotropic re-emission, we would expect a tighter post-\ac{ce} separation or higher fraction of mergers. We leave such exploration for future work. We additionally assume that the entire envelope will be lost in this second stage. Strictly speaking, the amount of envelope mass left on the core should depend on the size of the Roche lobe, but generally the envelope expands to large sizes unless it is stripped down to the last $\sim1/4$--$1/3$ of the radiative intershell \citep[]{iva11,vig22}. Indeed, some detailed 1D \ac{ce} simulations find that the donor detaches from its Roche lobe only when the remaining envelope mass is extremely small \citep[]{eld09,eld17,fra19,mar21}. The small amount of remaining envelope can strongly influence the observable features of the remaining star \citep[e.g.][]{hal13,got17}, its later evolution \citep[e.g.][]{lap20}, and the eventual supernova \citep[e.g.][]{des11,gil22}.  However, the retained envelope mass is small enough, except perhaps at low metallicities \citep[e.g.][]{got17}, that it should not significantly affect subsequent orbital evolution and final separation even if it leads to another mass transfer episode.

To demonstrate the predictions for the post-\ac{ce} orbits with our new formalism, we evolve a representative star with $M_\mathrm{d}=12~\msun$ through the \ac{rsg} phase using MESA. For simplicity, we ignore any mass loss that may occur via stellar winds or pulsations. We compute the core mass $M_\mathrm{core}$, convective envelope mass $M_\mathrm{conv}$, radiative intershell mass $M_\mathrm{int}$ and the binding energy of the convective envelope $E_\mathrm{conv}$ at each time. Inlists, extra subroutines and the output file used in this study are made available at \dataset[doi:10.5281/zenodo.7066430]{https://doi.org/10.5281/zenodo.7066430}. We use the above quantities to estimate the final post-\ac{ce} separation for various companion star masses. To provide a pre-\ac{ce} orbital separation, we assume that the donor has just filled its Roche lobe. The resulting separation is computed using the formula provided in \citet{egg83}. The true onset separation of the \ac{ce} phase is still heavily debated. In some cases it may be initiated via Darwin instability \citep[]{dar79}, especially for the more extreme mass ratio systems. In other cases the outflow through the second Lagrangian point may carry away angular momentum to drive the system into coalescence \citep[e.g.][]{mac18a}. When exactly the \ac{ce} commences and whether there are significant structural changes in the donor before the plunge-in is still unclear. In any case, the exact choice of the pre-\ac{ce} orbital separation has little influence on our demonstration as it only has a minor contribution in stage I through Eq.~(\ref{eq:alpha-formalism}). We set $\alpha_\mathrm{CE}=1$ and $\alpha_\mathrm{th}=1$ for stage I of our formalism and assume isotropic re-emission for stage II.

\section{Results}\label{sec:results}

\begin{figure}
 \centering
 \includegraphics[width=\linewidth]{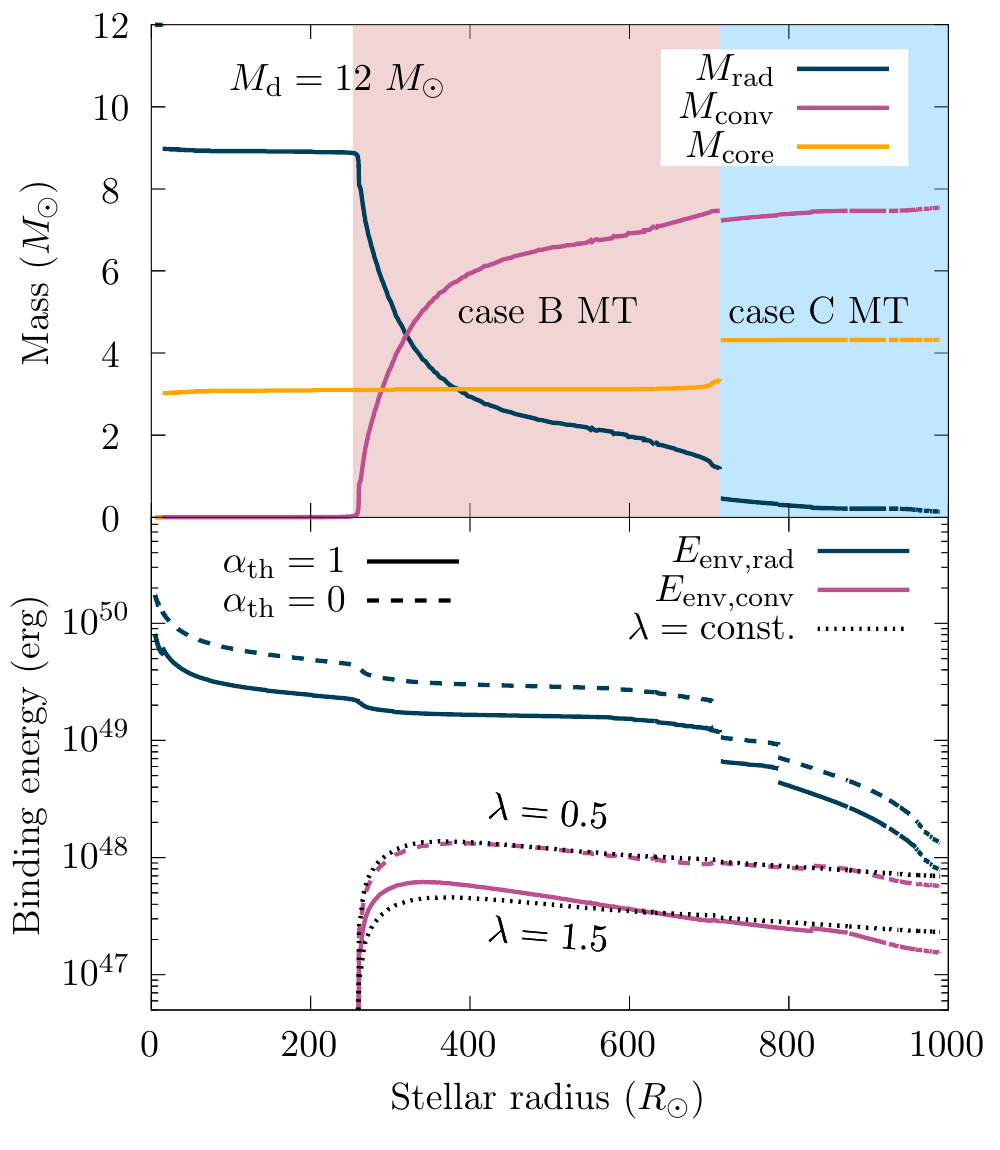}
 \caption{Key stellar quantities of a $12~\msun$ star as a function of the stellar radius. Values are plotted when the star reaches the given radius for the first time, which corresponds to when the donor starts transferring mass to the companion for the first time. \textit{(Upper panel)} Masses of the radiative intershell $M_\mathrm{rad}$, convective outer envelope $M_\mathrm{conv}$ and helium core $M_\mathrm{core}$. \textit{(Lower panel)} Binding energies of the radiative intershell $E_\mathrm{env,rad}$ (blue curves) and the convective outer envelope $E_\mathrm{env,conv}$ (magenta curves). The energies were computed with (solid) and without (dashed) internal energy. As a visual guide, we overplot simple estimates of the binding energy with fixed values of $\lambda$ (dotted curves). \label{fig:final_sep}}
\end{figure}

We show the evolution of key quantities for a representative model in Figure~\ref{fig:final_sep} with a $M_\mathrm{d}=12~\msun$ donor star. The upper panel shows how the convective envelope mass grows as the star evolves. Once the star reaches a certain threshold radius ($R_\mathrm{d}\sim250~\rsun$), a convective envelope develops and quickly grows up to when it reaches $R_\mathrm{d}\sim400~\rsun$ and then slowly grows up to its maximal value at the maximum radius ($R_\mathrm{d}\sim1000~\rsun$). The core mass stays almost constant while the star expands and crosses the Hertzsprung gap up to $R_\mathrm{d}\sim700~\rsun$. There is a steep jump in core mass beyond that radius, as the larger radii are only reached in later stages of the evolution when helium is depleted in the core (we do not display intermediate timesteps, when the star contracts and re-expands during core helium burning, as such stars will not generally initiate mass transfer in isolated binaries). If a \ac{ce} phase is initiated at radii within the pink shaded region, the interaction can be classified as case B mass transfer, whereas if it is initiated in the light blue region it can be classified as case C mass transfer. While the core mass and radiative intershell mass show a jump between case B and case C, the convective envelope mass seems to connect more smoothly.

The binding energies of the radiative intershell (blue curves) and the convective outer envelope (magenta curves) are shown in the lower panel. The radiative intershell has more than an order of magnitude higher binding energy throughout most of the evolution. Similar to the masses, the radiative intershell binding energy shows jumps between case B and case C \acp{ce}, but the convective envelope binding energy shows a smooth transition. The latter is almost perfectly consistent with having a constant value of $\lambda$ throughout, with $\lambda=0.5$ for the gravitational binding energy ($\alpha_\mathrm{th}=0$) and $\lambda\sim1.5$ when internal energy is included ($\alpha_\mathrm{th}=1$). This agreement was observed in all the masses we explored ($M_\mathrm{d}=8, 12, 20~\msun$), which could be a useful relation when generalizing our formalism.

\begin{figure*}
 \centering
 \includegraphics[width=0.8\linewidth]{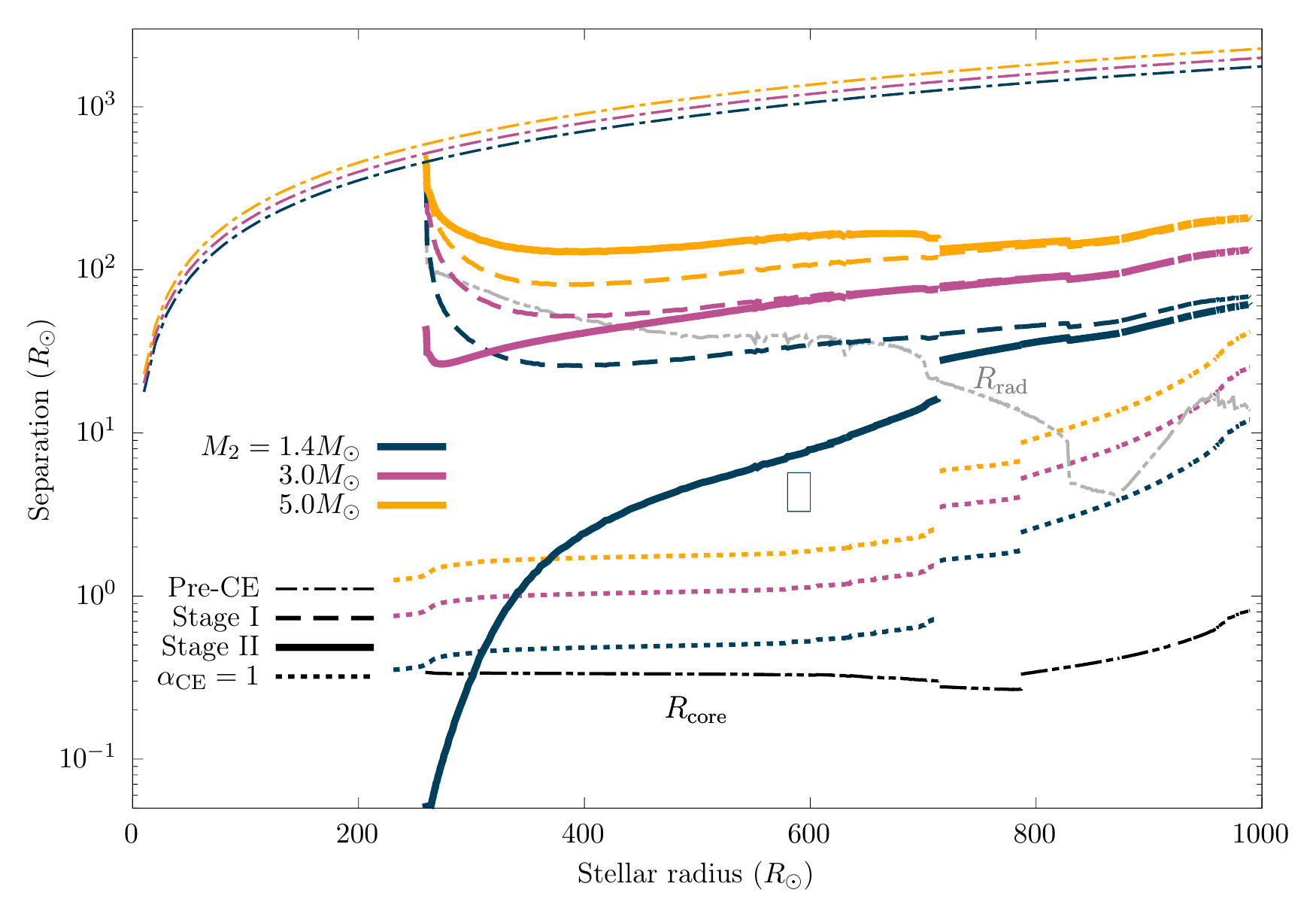}
 \caption{Post-\ac{ce} orbital separations predicted from our new formalism for 1.4, 3, and 5 $\msun$ companions as a function of the radius of the 12 $\msun$ donor when it first overflows its Roche lobe. Dot-dashed curves show the pre-\ac{ce} orbital separation, dashed curves show the orbital separations after stage I and solid curves show the final post-\ac{ce} separations. For comparison, we overplot the core radius $R_\mathrm{core}$ (black curve) and radius at the top of the intershell $R_\mathrm{rad}$ (grey curve), along with the post-\ac{ce} separations predicted by the classical $\alpha$-formalism with $\alpha_\mathrm{CE}=1$ (dotted curves). The blue rectangle indicates the \citet{fra19} result for a $12+1.4~\msun$ system.\label{fig:sep_comparison}}
\end{figure*}

Figure~\ref{fig:sep_comparison} displays the final separations predicted from our two-stage formalism. After stage I, the orbital separation shrinks roughly by an order of magnitude to $a\sim20$--$200~\rsun$ depending on the companion mass (dashed curves). The amount of shrinkage depends only weakly on the evolutionary stage of the donor when the \ac{ce} phase is initiated. In some cases when the companion mass is small, the separation after stage I becomes smaller than the outer radius of the radiative intershell $R_\mathrm{rad}$. It is not trivial what should happen in such cases. There is little mass in the outer $\sim2/3$ of the radiative intershell, so it may still be possible to smoothly transition into a sufficiently stable mass transfer phase if it only ducks into the radiative intershell by a small amount. Even if it dives deep into the radiative intershell, it may still be possible to eject the envelope via adiabatic processes as explored by \citet{law20}. Conversely, the dynamical friction within the radiative intershell may be strong enough to cause the companion to merge with the donor core. Or the transferred mass may develop into a contact binary configuration in which the angular momentum is taken away from the $L_2$ point. In this paper, we assume that it smoothly transitions to stage II in all cases. However, it may be sensible in future studies to consider alternative treatments for the cases where the companion dives into the radiative intershell after stage I.

The following stage II creates a dramatic difference in the final separations depending on the companion mass. The orbit shrinks with stable mass transfer when the companion mass is smaller than the core mass ($M_2<M_\mathrm{core}$), whereas the orbit expands when the companion is heavier than the core ($M_2>M_\mathrm{core}$). When the core is similar to the companion mass ($M_2\sim M_\mathrm{core}$), the separation hardly changes after stage II. There is also little change for all case C \acp{ce}, because the intershell mass is much smaller at these stages. We note that our separation predicted for a $12+1.4~\msun$ system with a $R_\mathrm{d}=600~\rsun$ donor is within a factor $\lesssim1.5$ of the results obtained from 1D \ac{ce} simulations by \citet{fra19} despite the drastically different approaches.

\begin{figure}
 \centering
 \includegraphics[width=\linewidth]{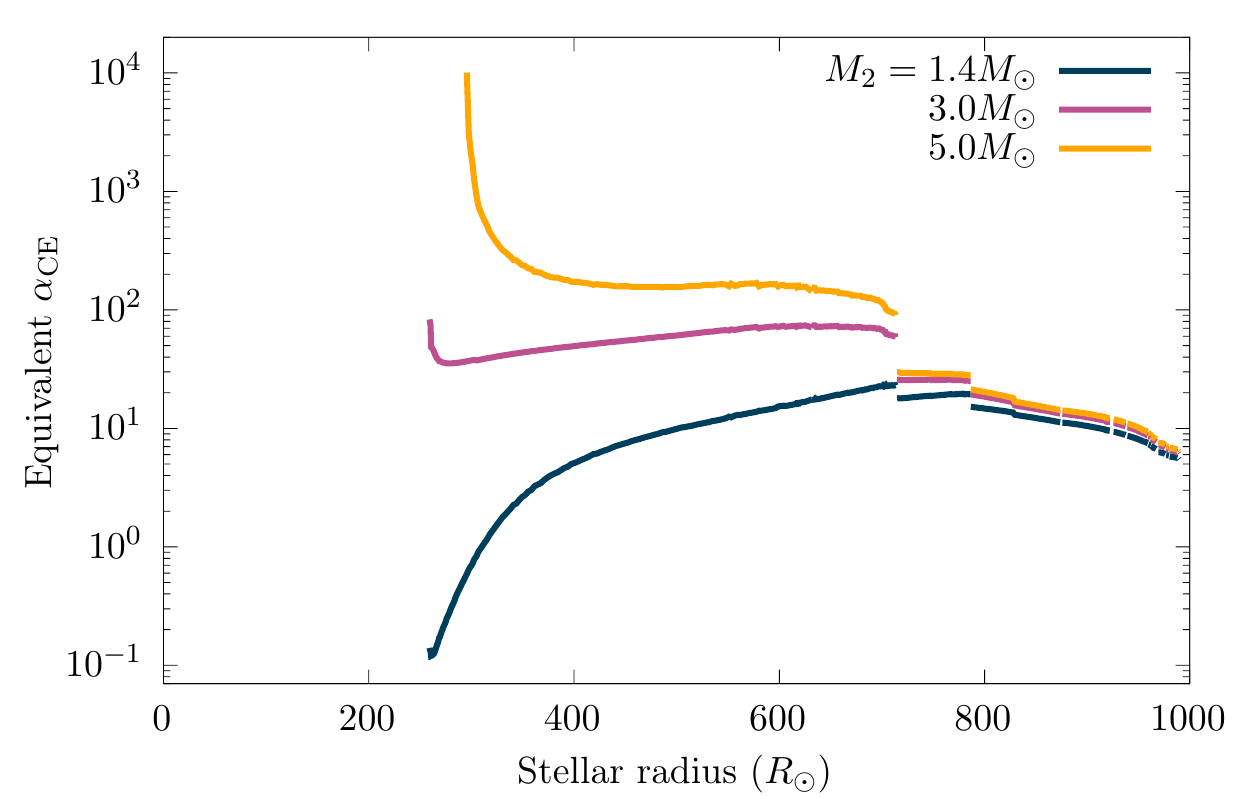}
 \caption{The value of {\ace} required to reproduce the final separations predicted from our new formalism.\label{fig:equivalent_alpha}}
\end{figure}

Our formalism generally predicts rather large post-\ac{ce} separations, ranging from a few tens to a few hundred $\rsun$, except for accretors with the mass of a neutron star, which can yield much smaller separations. In fact, most systems seem to avoid either a merger with the core or a Roche lobe overflow from a companion.  This is in stark contrast with the post-\ac{ce} separations predicted from the classical $\alpha$-formalism (dotted curves), which are orders of magnitude smaller throughout most of the parameter space when we set $\alpha_\mathrm{CE}=1$.

In Figure~\ref{fig:equivalent_alpha}, we plot the values of {\ace} required in the classical $\alpha$-formalism to reproduce the final separations from our new formalism. In most cases, our new formalism is equivalent to setting {\ace} to very high values ($\alpha_\mathrm{CE}>10$). The only exception is for early case B \acp{ce} with lower-mass companions. Such high values of {\ace} have not typically been considered in binary population synthesis studies. There are no direct observational constraints on {\ace} for massive stars, but similar ranges of {\ace} are often assumed as those inferred from observations and simulations of low-mass stars, which are generally low \citep[]{iac19}.  We note that the BPASS code \citep[]{eld09,eld17} takes an original approach to \ac{ce} phases and \citet[]{eld17} report that they have equivalent $\alpha_\mathrm{CE}\lambda$ values ranging between 2 to 100, with the majority between 2 to 30. Assuming that their $\lambda$ values are similar to the typical values including the intershell ($\lambda\sim0.1$--$0.5$), that corresponds to $\alpha_\mathrm{CE}\sim4$--$300$, which is similar to what we predict in our new formalism. However, they also report that the \ac{ce} efficiency is highest when the mass ratio is smallest, an opposite trend from what we find. Thus the similarity may be coincidental.

There is a narrow range with the larger companion mass ($M_2=5~\msun$) where we predict a final separation which is impossible to achieve in the $\alpha$-formalism no matter what value of $\alpha_\mathrm{CE}(>0)$ is used (when $a_f/a_i\geq M_\mathrm{core}/M_\mathrm{d}$). This only occurs in regions when the convective envelope is still shallow and therefore the whole process is more like fully stable mass transfer with mass ratio reversal than a \ac{ce} phase. We caution that in these regimes where the star is still close to being fully radiative, our assumption of stable mass transfer may break down due to the so-called delayed dynamical instability \citep[]{hje87}.

Most importantly, our new formalism has highly variable equivalent \ace's depending on the evolutionary stage and companion masses (see Figure~\ref{fig:equivalent_alpha}). In the classical $\alpha$-formalism, the post-\ac{ce} separation scales linearly with the companion mass (Figure~\ref{fig:sep_comparison}). This suggests that using universal values for {\ace} in rapid binary population synthesis studies will not achieve even qualitatively accurate predictions.

Our illustrative examples used a particular set of MESA models where we assumed mostly default settings with $Z=0.014$ for the metallicity, along with a convective step overshooting parameter of $\alpha_\mathrm{ov}=0.33$ and a semiconvection parameter of $\alpha_\mathrm{sc}=0.01$. Other assumptions can yield markedly different relationships between convective envelope binding energies, radiative intershell masses, and stellar radii. These variations include the treatment of mixing processes \citep{sch19,kle21}, and the effect of previous accretion onto the donor \citep[]{ren22}. While the specific \ac{ce} outcomes will depend on the details of stellar evolution models, the fundamental physical considerations that motivate the two-stage approach proposed here still hold.

\section{Conclusion and discussion}\label{sec:conclusion}

We propose a new formalism to predict the outcomes of \ac{ce} phases with massive star donors. In our new formalism, we split the whole \ac{ce} process into two stages: (I) dynamical inspiral through the outer convective part of the envelope, and (II) thermal timescale mass transfer of the inner radiative part of the envelope. We treat stage I as an adiabatic process and apply the commonly used $\alpha$-formalism (or energy formalism). Stage II is treated as a stable mass transfer, focusing on the angular momentum evolution of the system. The combined outcome gives the final post-\ac{ce} separation.

We demonstrated the predictions for the final separations with our formalism with some fiducial choices of parameters ($\alpha_\mathrm{CE}=1$ for stage I and isotropic re-emission for stage II). Overall, the predicted post-\ac{ce} orbital separations seem to be rather wide compared to what would be predicted from classical energy formalisms. There is a qualitative difference depending on whether the companion mass is larger or smaller than the donor core mass. When the companion is lighter than the core, the orbit can shrink substantially down to values similar to those predicted by traditional formalisms. On the other hand when the companion is heavier than the core, the orbit can widen after the initial plunge-in, ending up at large orbital separations.

The generally wide post-\ac{ce} separations can have broad implications. One example is that we may predict a high fraction of stripped-envelope supernova progenitors to have low-mass companions, which previously would have been considered unable to survive \acp{ce}. In a supernova called SN2006jc, we estimated from the current companion photometry that the progenitor system experienced a \ac{ce} phase with a $M_\mathrm{d}\sim12~\msun$ donor and $M_2\sim3$--$4~\msun$ companion, ending up with a $a\sim40~\rsun$ separation \citep[]{sun20,oga21}. Our new formalism predicts $a\sim40$--$100~\rsun$ for $M_2=3~\msun$, which is roughly in agreement with this observational constraint. Another example may be iPTF13bvn \citep[]{cao13}. Based on its classification as a type Ib supernova and on the progenitor radius inferred from pre-supernova photometry, we argued that it should have experienced a \ac{ce} phase with $\alpha_\mathrm{CE}\gtrsim20$ \citep[]{RH17a}. Such high efficiencies are easily achievable in our new formalism, implying that iPTF13bvn may have a low-mass companion star. Many other stripped-envelope supernova progenitors may also have dim low-mass companions, which may explain the lack of companion detections. 

The increase in the post-\ac{ce} separation predicted by our two-stage formalism may explain how binaries with relatively extreme mass ratios may survive mass transfer, addressing a puzzle in the formation of low-mass X-ray binaries \citep[]{pod00,pod02}.  The wide separations predicted for massive companions also indicate that \acp{ce} are unlikely to contribute significantly to the formation of binary black holes as gravitational-wave sources, since extreme mass ratios are unlikely when transferring mass onto a black hole formed from the primary and $30 \msun$ black hole binaries with separations above $\sim 50 \rsun$ will not merge in the age of Universe.  On the other hand, our formalism can reproduce very tight Galactic double neutron star systems observed as radio pulsars \citep[e.g.][]{hul75}, provided a \ac{ce} stage initiated by an early Hertzsprung-gap donor is survivable.

One of the main features of our formalism is that the \ac{ce} efficiency has a strong dependence on the system parameters such as the evolutionary stage of the donor and the companion mass. When translated to the traditional $\alpha$-formalism, our final separations correspond to an extremely wide range of equivalent \ace{} values from $\sim0.1$ to $>10^4$. This should have a significant impact on the results of population synthesis studies, which currently typically employ a universal value of {\ace}. Frequently, the uncertainties are explored by varying the value of the universal \ace{}, but that cannot reproduce the qualitative sensitivity to the donor's evolutionary stage and companion mass.  We leave it for future work to explore how the predictions of the populations of various objects such as binary black hole and binary neutron star mergers, X-ray binaries, double pulsars and stripped-envelope supernovae will be impacted with our new formalism.

We stress that we are not trying to establish a complete model, but rather a physically-motivated framework as a basis for further research. Many of the assumptions going into this formalism should be examined in future work. Some other intricacies such as pre-\ac{ce} evolution, onset of \ac{ce} phases, envelope fallback, etc have been ignored to maintain simplicity for our formalism. However, we believe that our model captures the key features that distinguish massive star donors from low-mass donors, and provides a useful guideline for how to treat \acp{ce} in population synthesis studies.

\begin{acknowledgments}
The authors thank Orsola De Marco, Mike Lau, and Daniel Price for discussions, and Lewis Picker for comments on the manuscript. This work was performed on the OzSTAR national facility at Swinburne University of Technology. The OzSTAR program receives funding in part from the Astronomy National Collaborative Research Infrastructure Strategy (NCRIS) allocation provided by the Australian Government.
IM is a recipient of the Australian Research Council Future Fellowship FT190100574.
\end{acknowledgments}

%



\software{MESA \citep{MESA1,MESA2,MESA3,MESA4,MESA5},
          MESA SDK \citep{mesasdk},
          gnuplot \citep{Gnuplot}
          }





\bibliographystyle{aasjournal}

\begin{thebibliography}{}
\expandafter\ifx\csname natexlab\endcsname\relax\def\natexlab#1{#1}\fi
\providecommand{\url}[1]{\href{#1}{#1}}
\providecommand{\dodoi}[1]{doi:~\href{http://doi.org/#1}{\nolinkurl{#1}}}
\providecommand{\doeprint}[1]{\href{http://ascl.net/#1}{\nolinkurl{http://ascl.net/#1}}}
\providecommand{\doarXiv}[1]{\href{https://arxiv.org/abs/#1}{\nolinkurl{https://arxiv.org/abs/#1}}}

\bibitem[{{Cao} {et~al.}(2013){Cao}, {Kasliwal}, {Arcavi}, {Horesh}, {Hancock},
  {Valenti}, {Cenko}, {Kulkarni}, {Gal-Yam}, {Gorbikov}, {Ofek}, {Sand},
  {Yaron}, {Graham}, {Silverman}, {Wheeler}, {Marion}, {Walker}, {Mazzali},
  {Howell}, {Li}, {Kong}, {Bloom}, {Nugent}, {Surace}, {Masci}, {Carpenter},
  {Degenaar}, \& {Gelino}}]{cao13}
{Cao}, Y., {Kasliwal}, M.~M., {Arcavi}, I., {et~al.} 2013, \apjl, 775, L7,
  \dodoi{10.1088/2041-8205/775/1/L7}

\bibitem[{{Chamandy} {et~al.}(2018){Chamandy}, {Frank}, {Blackman},
  {Carroll-Nellenback}, {Liu}, {Tu}, {Nordhaus}, {Chen}, \& {Peng}}]{cha18}
{Chamandy}, L., {Frank}, A., {Blackman}, E.~G., {et~al.} 2018, \mnras, 480,
  1898, \dodoi{10.1093/mnras/sty1950}

\bibitem[{{Clayton} {et~al.}(2017){Clayton}, {Podsiadlowski}, {Ivanova}, \&
  {Justham}}]{cla17}
{Clayton}, M., {Podsiadlowski}, {\relax Ph}., {Ivanova}, N., \& {Justham}, S.
  2017, \mnras, 470, 1788, \dodoi{10.1093/mnras/stx1290}

\bibitem[{{Darwin}(1879)}]{dar79}
{Darwin}, G.~H. 1879, Proc. R. Soc. Lond., 29, 168,
  \dodoi{10.1098/rspl.1879.0028}

\bibitem[{{de Kool}(1990)}]{dek90}
{de Kool}, M. 1990, \apj, 358, 189, \dodoi{10.1086/168974}

\bibitem[{{Dessart} {et~al.}(2011){Dessart}, {Hillier}, {Livne}, {Yoon},
  {Woosley}, {Waldman}, \& {Langer}}]{des11}
{Dessart}, L., {Hillier}, D.~J., {Livne}, E., {et~al.} 2011, \mnras, 414, 2985,
  \dodoi{10.1111/j.1365-2966.2011.18598.x}

\bibitem[{{Dewi} \& {Tauris}(2000)}]{dew00}
{Dewi}, J.~D.~M., \& {Tauris}, T.~M. 2000, \aap, 360, 1043.
\newblock \doarXiv{astro-ph/0007034}

\bibitem[{{Eggleton}(1983)}]{egg83}
{Eggleton}, P.~P. 1983, \apj, 268, 368, \dodoi{10.1086/160960}

\bibitem[{{Eldridge} \& {Stanway}(2009)}]{eld09}
{Eldridge}, J.~J., \& {Stanway}, E.~R. 2009, \mnras, 400, 1019,
  \dodoi{10.1111/j.1365-2966.2009.15514.x}

\bibitem[{{Eldridge} {et~al.}(2017){Eldridge}, {Stanway}, {Xiao}, {McClelland},
  {Taylor}, {Ng}, {Greis}, \& {Bray}}]{eld17}
{Eldridge}, J.~J., {Stanway}, E.~R., {Xiao}, L., {et~al.} 2017, \pasa, 34,
  e058, \dodoi{10.1017/pasa.2017.51}

\bibitem[{{Fragos} {et~al.}(2019){Fragos}, {Andrews}, {Ramirez-Ruiz}, {Meynet},
  {Kalogera}, {Taam}, \& {Zezas}}]{fra19}
{Fragos}, T., {Andrews}, J.~J., {Ramirez-Ruiz}, E., {et~al.} 2019, \apjl, 883,
  L45, \dodoi{10.3847/2041-8213/ab40d1}

\bibitem[{{Fryer} {et~al.}(1996){Fryer}, {Benz}, \& {Herant}}]{fry96}
{Fryer}, C.~L., {Benz}, W., \& {Herant}, M. 1996, \apj, 460, 801,
  \dodoi{10.1086/177011}

\bibitem[{{Gilkis} \& {Arcavi}(2022)}]{gil22}
{Gilkis}, A., \& {Arcavi}, I. 2022, \mnras, 511, 691,
  \dodoi{10.1093/mnras/stac088}

\bibitem[{{Glanz} \& {Perets}(2021)}]{gla21}
{Glanz}, H., \& {Perets}, H.~B. 2021, \mnras, 507, 2659,
  \dodoi{10.1093/mnras/stab2291}

\bibitem[{{Gonzalez-Bolivar} {et~al.}(2022){Gonzalez-Bolivar}, {De Marco},
  {Lau}, {Hirai}, \& {Price}}]{gon22}
{Gonzalez-Bolivar}, M., {De Marco}, O., {Lau}, M. Y.~M., {Hirai}, R., \&
  {Price}, D.~J. 2022, arXiv e-prints, arXiv:2205.09749.
\newblock \doarXiv{2205.09749}

\bibitem[{{G{\"o}tberg} {et~al.}(2017){G{\"o}tberg}, {de Mink}, \&
  {Groh}}]{got17}
{G{\"o}tberg}, Y., {de Mink}, S.~E., \& {Groh}, J.~H. 2017, \aap, 608, A11,
  \dodoi{10.1051/0004-6361/201730472}

\bibitem[{{Hall} {et~al.}(2013){Hall}, {Tout}, {Izzard}, \& {Keller}}]{hal13}
{Hall}, P.~D., {Tout}, C.~A., {Izzard}, R.~G., \& {Keller}, D. 2013, \mnras,
  435, 2048, \dodoi{10.1093/mnras/stt1422}

\bibitem[{{Han} {et~al.}(1995){Han}, {Podsiadlowski}, \& {Eggleton}}]{han95}
{Han}, Z., {Podsiadlowski}, {\relax Ph}., \& {Eggleton}, P.~P. 1995, \mnras,
  272, 800, \dodoi{10.1093/mnras/272.4.800}

\bibitem[{{Hirai}(2017)}]{RH17a}
{Hirai}, R. 2017, \mnras, 466, 3775, \dodoi{10.1093/mnras/stw3321}

\bibitem[{{Hjellming} \& {Webbink}(1987)}]{hje87}
{Hjellming}, M.~S., \& {Webbink}, R.~F. 1987, \apj, 318, 794,
  \dodoi{10.1086/165412}

\bibitem[{{Hulse} \& {Taylor}(1975)}]{hul75}
{Hulse}, R.~A., \& {Taylor}, J.~H. 1975, \apjl, 195, L51,
  \dodoi{10.1086/181708}

\bibitem[{{Iaconi} \& {De Marco}(2019)}]{iac19}
{Iaconi}, R., \& {De Marco}, O. 2019, \mnras, 490, 2550,
  \dodoi{10.1093/mnras/stz2756}

\bibitem[{{Iaconi} {et~al.}(2017){Iaconi}, {Reichardt}, {Staff}, {De Marco},
  {Passy}, {Price}, {Wurster}, \& {Herwig}}]{iac17}
{Iaconi}, R., {Reichardt}, T., {Staff}, J., {et~al.} 2017, \mnras, 464, 4028,
  \dodoi{10.1093/mnras/stw2377}

\bibitem[{{Iben} \& {Tutukov}(1985)}]{ibe85}
{Iben}, I., J., \& {Tutukov}, A.~V. 1985, \apjs, 58, 661,
  \dodoi{10.1086/191054}

\bibitem[{{Ivanova}(2011)}]{iva11}
{Ivanova}, N. 2011, \apj, 730, 76, \dodoi{10.1088/0004-637X/730/2/76}

\bibitem[{{Ivanova} {et~al.}(2015){Ivanova}, {Justham}, \&
  {Podsiadlowski}}]{iva15}
{Ivanova}, N., {Justham}, S., \& {Podsiadlowski}, {\relax Ph}. 2015, \mnras,
  447, 2181, \dodoi{10.1093/mnras/stu2582}

\bibitem[{{Ivanova} {et~al.}(2020){Ivanova}, {Justham}, \& {Ricker}}]{iva20}
{Ivanova}, N., {Justham}, S., \& {Ricker}, P. 2020, {Common Envelope
  Evolution}, \dodoi{10.1088/2514-3433/abb6f0}

\bibitem[{{Ivanova} \& {Nandez}(2016)}]{iva16}
{Ivanova}, N., \& {Nandez}, J.~L.~A. 2016, \mnras, 462, 362,
  \dodoi{10.1093/mnras/stw1676}

\bibitem[{{Ivanova} {et~al.}(2013){Ivanova}, {Justham}, {Chen}, {De Marco},
  {Fryer}, {Gaburov}, {Ge}, {Glebbeek}, {Han}, {Li}, {Lu}, {Marsh},
  {Podsiadlowski}, {Potter}, {Soker}, {Taam}, {Tauris}, {van den Heuvel}, \&
  {Webbink}}]{iva13}
{Ivanova}, N., {Justham}, S., {Chen}, X., {et~al.} 2013, \aapr, 21, 59,
  \dodoi{10.1007/s00159-013-0059-2}

\bibitem[{{Klencki} {et~al.}(2021){Klencki}, {Nelemans}, {Istrate}, \&
  {Chruslinska}}]{kle21}
{Klencki}, J., {Nelemans}, G., {Istrate}, A.~G., \& {Chruslinska}, M. 2021,
  \aap, 645, A54, \dodoi{10.1051/0004-6361/202038707}

\bibitem[{{Laplace} {et~al.}(2020){Laplace}, {G{\"o}tberg}, {de Mink},
  {Justham}, \& {Farmer}}]{lap20}
{Laplace}, E., {G{\"o}tberg}, Y., {de Mink}, S.~E., {Justham}, S., \& {Farmer},
  R. 2020, \aap, 637, A6, \dodoi{10.1051/0004-6361/201937300}

\bibitem[{{Lau} {et~al.}(2022{\natexlab{a}}){Lau}, {Hirai},
  {Gonz{\'a}lez-Bol{\'\i}var}, {Price}, {De Marco}, \& {Mandel}}]{lau22a}
{Lau}, M. Y.~M., {Hirai}, R., {Gonz{\'a}lez-Bol{\'\i}var}, M., {et~al.}
  2022{\natexlab{a}}, \mnras, 512, 5462, \dodoi{10.1093/mnras/stac049}

\bibitem[{{Lau} {et~al.}(2022{\natexlab{b}}){Lau}, {Hirai}, {Price}, \&
  {Mandel}}]{lau22b}
{Lau}, M. Y.~M., {Hirai}, R., {Price}, D.~J., \& {Mandel}, I.
  2022{\natexlab{b}}, arXiv e-prints, arXiv:2206.06411.
\newblock \doarXiv{2206.06411}

\bibitem[{{Law-Smith} {et~al.}(2020){Law-Smith}, {Everson}, {Ramirez-Ruiz}, {de
  Mink}, {van Son}, {G{\"o}tberg}, {Zellmann}, {Vigna-G{\'o}mez}, {Renzo},
  {Wu}, {Schr{\o}der}, {Foley}, \& {Hutchinson-Smith}}]{law20}
{Law-Smith}, J. A.~P., {Everson}, R.~W., {Ramirez-Ruiz}, E., {et~al.} 2020,
  arXiv e-prints, arXiv:2011.06630.
\newblock \doarXiv{2011.06630}

\bibitem[{{L{\'o}pez-C{\'a}mara} {et~al.}(2019){L{\'o}pez-C{\'a}mara}, {De
  Colle}, \& {Moreno M{\'e}ndez}}]{lop19}
{L{\'o}pez-C{\'a}mara}, D., {De Colle}, F., \& {Moreno M{\'e}ndez}, E. 2019,
  \mnras, 482, 3646, \dodoi{10.1093/mnras/sty2959}

\bibitem[{{MacLeod} {et~al.}(2018{\natexlab{a}}){MacLeod}, {Ostriker}, \&
  {Stone}}]{mac18b}
{MacLeod}, M., {Ostriker}, E.~C., \& {Stone}, J.~M. 2018{\natexlab{a}}, \apj,
  868, 136, \dodoi{10.3847/1538-4357/aae9eb}

\bibitem[{{MacLeod} {et~al.}(2018{\natexlab{b}}){MacLeod}, {Ostriker}, \&
  {Stone}}]{mac18a}
---. 2018{\natexlab{b}}, \apj, 863, 5, \dodoi{10.3847/1538-4357/aacf08}

\bibitem[{{Marchant} {et~al.}(2021){Marchant}, {Pappas}, {Gallegos-Garcia},
  {Berry}, {Taam}, {Kalogera}, \& {Podsiadlowski}}]{mar21}
{Marchant}, P., {Pappas}, K. M.~W., {Gallegos-Garcia}, M., {et~al.} 2021, \aap,
  650, A107, \dodoi{10.1051/0004-6361/202039992}

\bibitem[{{Moreno} {et~al.}(2021){Moreno}, {Schneider}, {Roepke}, {Ohlmann},
  {Pakmor}, {Podsiadlowski}, \& {Sand}}]{mor21}
{Moreno}, M.~M., {Schneider}, F. R.~N., {Roepke}, F.~K., {et~al.} 2021, arXiv
  e-prints, arXiv:2111.12112.
\newblock \doarXiv{2111.12112}

\bibitem[{{Nelemans} \& {Tout}(2005)}]{nel05}
{Nelemans}, G., \& {Tout}, C.~A. 2005, \mnras, 356, 753,
  \dodoi{10.1111/j.1365-2966.2004.08496.x}

\bibitem[{{Nelemans} {et~al.}(2000){Nelemans}, {Verbunt}, {Yungelson}, \&
  {Portegies Zwart}}]{nel00}
{Nelemans}, G., {Verbunt}, F., {Yungelson}, L.~R., \& {Portegies Zwart}, S.~F.
  2000, \aap, 360, 1011.
\newblock \doarXiv{astro-ph/0006216}

\bibitem[{{Ogata} {et~al.}(2021){Ogata}, {Hirai}, \& {Hijikawa}}]{oga21}
{Ogata}, M., {Hirai}, R., \& {Hijikawa}, K. 2021, \mnras, 505, 2485,
  \dodoi{10.1093/mnras/stab1439}

\bibitem[{{Ohlmann} {et~al.}(2016){Ohlmann}, {R{\"o}pke}, {Pakmor}, {Springel},
  \& {M{\"u}ller}}]{ohl16}
{Ohlmann}, S.~T., {R{\"o}pke}, F.~K., {Pakmor}, R., {Springel}, V., \&
  {M{\"u}ller}, E. 2016, \mnras, 462, L121, \dodoi{10.1093/mnrasl/slw144}

\bibitem[{{Paczynski}(1976)}]{pac76}
{Paczynski}, B. 1976, in Structure and Evolution of Close Binary Systems, ed.
  P.~{Eggleton}, S.~{Mitton}, \& J.~{Whelan}, Vol.~73, 75

\bibitem[{{Passy} {et~al.}(2012){Passy}, {De Marco}, {Fryer}, {Herwig},
  {Diehl}, {Oishi}, {Mac Low}, {Bryan}, \& {Rockefeller}}]{pas12}
{Passy}, J.-C., {De Marco}, O., {Fryer}, C.~L., {et~al.} 2012, \apj, 744, 52,
  \dodoi{10.1088/0004-637X/744/1/52}

\bibitem[{{Paxton} {et~al.}(2011){Paxton}, {Bildsten}, {Dotter}, {Herwig},
  {Lesaffre}, \& {Timmes}}]{MESA1}
{Paxton}, B., {Bildsten}, L., {Dotter}, A., {et~al.} 2011, \apjs, 192, 3,
  \dodoi{10.1088/0067-0049/192/1/3}

\bibitem[{{Paxton} {et~al.}(2013){Paxton}, {Cantiello}, {Arras}, {Bildsten},
  {Brown}, {Dotter}, {Mankovich}, {Montgomery}, {Stello}, {Timmes}, \&
  {Townsend}}]{MESA2}
{Paxton}, B., {Cantiello}, M., {Arras}, P., {et~al.} 2013, \apjs, 208, 4,
  \dodoi{10.1088/0067-0049/208/1/4}

\bibitem[{{Paxton} {et~al.}(2015){Paxton}, {Marchant}, {Schwab}, {Bauer},
  {Bildsten}, {Cantiello}, {Dessart}, {Farmer}, {Hu}, {Langer}, {Townsend},
  {Townsley}, \& {Timmes}}]{MESA3}
{Paxton}, B., {Marchant}, P., {Schwab}, J., {et~al.} 2015, \apjs, 220, 15,
  \dodoi{10.1088/0067-0049/220/1/15}

\bibitem[{{Paxton} {et~al.}(2018){Paxton}, {Schwab}, {Bauer}, {Bildsten},
  {Blinnikov}, {Duffell}, {Farmer}, {Goldberg}, {Marchant}, {Sorokina},
  {Thoul}, {Townsend}, \& {Timmes}}]{MESA4}
{Paxton}, B., {Schwab}, J., {Bauer}, E.~B., {et~al.} 2018, \apjs, 234, 34,
  \dodoi{10.3847/1538-4365/aaa5a8}

\bibitem[{{Paxton} {et~al.}(2019){Paxton}, {Smolec}, {Schwab}, {Gautschy},
  {Bildsten}, {Cantiello}, {Dotter}, {Farmer}, {Goldberg}, {Jermyn}, {Kanbur},
  {Marchant}, {Thoul}, {Townsend}, {Wolf}, {Zhang}, \& {Timmes}}]{MESA5}
{Paxton}, B., {Smolec}, R., {Schwab}, J., {et~al.} 2019, \apjs, 243, 10,
  \dodoi{10.3847/1538-4365/ab2241}

\bibitem[{{Podsiadlowski} \& {Rappaport}(2000)}]{pod00}
{Podsiadlowski}, {\relax Ph}., \& {Rappaport}, S. 2000, \apj, 529, 946,
  \dodoi{10.1086/308323}

\bibitem[{{Podsiadlowski} {et~al.}(2002){Podsiadlowski}, {Rappaport}, \&
  {Pfahl}}]{pod02}
{Podsiadlowski}, {\relax Ph}., {Rappaport}, S., \& {Pfahl}, E.~D. 2002, \apj,
  565, 1107, \dodoi{10.1086/324686}

\bibitem[{{Politano}(2004)}]{pol04}
{Politano}, M. 2004, \apj, 604, 817, \dodoi{10.1086/381958}

\bibitem[{{Politano}(2021)}]{pol21}
---. 2021, \aap, 648, L6, \dodoi{10.1051/0004-6361/202140442}

\bibitem[{{Politano} \& {Weiler}(2007)}]{pol07}
{Politano}, M., \& {Weiler}, K.~P. 2007, \apj, 665, 663, \dodoi{10.1086/518997}

\bibitem[{{Postnov} \& {Yungelson}(2014)}]{pos14}
{Postnov}, K.~A., \& {Yungelson}, L.~R. 2014, Living Reviews in Relativity, 17,
  3, \dodoi{10.12942/lrr-2014-3}

\bibitem[{{Rasio} \& {Livio}(1996)}]{ras96}
{Rasio}, F.~A., \& {Livio}, M. 1996, \apj, 471, 366, \dodoi{10.1086/177975}

\bibitem[{{Renzo} {et~al.}(2022){Renzo}, {Zapartas}, {Justham}, {Breivik},
  {Lau}, {Farmer}, {Cantiello}, \& {Metzger}}]{ren22}
{Renzo}, M., {Zapartas}, E., {Justham}, S., {et~al.} 2022, arXiv e-prints,
  arXiv:2206.15338.
\newblock \doarXiv{2206.15338}

\bibitem[{{Ricker} \& {Taam}(2008)}]{ric08}
{Ricker}, P.~M., \& {Taam}, R.~E. 2008, \apjl, 672, L41, \dodoi{10.1086/526343}

\bibitem[{{Schootemeijer} {et~al.}(2019){Schootemeijer}, {Langer}, {Grin}, \&
  {Wang}}]{sch19}
{Schootemeijer}, A., {Langer}, N., {Grin}, N.~J., \& {Wang}, C. 2019, \aap,
  625, A132, \dodoi{10.1051/0004-6361/201935046}

\bibitem[{{Soberman} {et~al.}(1997){Soberman}, {Phinney}, \& {van den
  Heuvel}}]{sob97}
{Soberman}, G.~E., {Phinney}, E.~S., \& {van den Heuvel}, E.~P.~J. 1997, \aap,
  327, 620.
\newblock \doarXiv{astro-ph/9703016}

\bibitem[{{Soker}(2004)}]{sok04}
{Soker}, N. 2004, \na, 9, 399, \dodoi{10.1016/j.newast.2004.01.004}

\bibitem[{{Sun} {et~al.}(2020){Sun}, {Maund}, {Hirai}, {Crowther}, \&
  {Podsiadlowski}}]{sun20}
{Sun}, N.-C., {Maund}, J.~R., {Hirai}, R., {Crowther}, P.~A., \&
  {Podsiadlowski}, {\relax Ph}. 2020, \mnras, 491, 6000,
  \dodoi{10.1093/mnras/stz3431}

\bibitem[{Townsend(2021)}]{mesasdk}
Townsend, R. 2021, MESA SDK for Linux, 21.4.1,  Zenodo,
  \dodoi{10.5281/zenodo.5802444}

\bibitem[{{Tutukov} \& {Yungelson}(1979)}]{tut79}
{Tutukov}, A., \& {Yungelson}, L. 1979, in Mass Loss and Evolution of O-Type
  Stars, ed. P.~S. {Conti} \& C.~W.~H. {De Loore}, Vol.~83, 401--406

\bibitem[{{van den Heuvel}(1976)}]{van76}
{van den Heuvel}, E.~P.~J. 1976, in Structure and Evolution of Close Binary
  Systems, ed. P.~{Eggleton}, S.~{Mitton}, \& J.~{Whelan}, Vol.~73, 35

\bibitem[{{Vigna-G{\'o}mez} {et~al.}(2022){Vigna-G{\'o}mez}, {Wassink},
  {Klencki}, {Istrate}, {Nelemans}, \& {Mandel}}]{vig22}
{Vigna-G{\'o}mez}, A., {Wassink}, M., {Klencki}, J., {et~al.} 2022, \mnras,
  511, 2326, \dodoi{10.1093/mnras/stac237}

\bibitem[{{Webbink}(1984)}]{web84}
{Webbink}, R.~F. 1984, \apj, 277, 355, \dodoi{10.1086/161701}

\bibitem[{{Webbink}(2008)}]{web08}
{Webbink}, R.~F. 2008, in Astrophysics and Space Science Library, Vol. 352,
  Astrophysics and Space Science Library, ed. E.~F. {Milone}, D.~A. {Leahy}, \&
  D.~W. {Hobill}, 233, \dodoi{10.1007/978-1-4020-6544-6_13}

\bibitem[{Williams {et~al.}(2021)Williams, Kelley, \& {many others}}]{Gnuplot}
Williams, T., Kelley, C., \& {many others}. 2021, Gnuplot 5.4: an interactive
  plotting program, \url{http://www.gnuplot.info}

\bibitem[{{Wilson} \& {Nordhaus}(2022)}]{wil22}
{Wilson}, E.~C., \& {Nordhaus}, J. 2022, arXiv e-prints, arXiv:2203.06091.
\newblock \doarXiv{2203.06091}

\bibitem[{{Xu} \& {Li}(2010{\natexlab{a}})}]{xu10a}
{Xu}, X.-J., \& {Li}, X.-D. 2010{\natexlab{a}}, \apj, 716, 114,
  \dodoi{10.1088/0004-637X/716/1/114}

\bibitem[{{Xu} \& {Li}(2010{\natexlab{b}})}]{xu10b}
---. 2010{\natexlab{b}}, \apj, 722, 1985, \dodoi{10.1088/0004-637X/722/2/1985}

\end{thebibliography}


\end{CJK*}
\end{document}